\documentstyle[12pt]{article}

\newcommand{\ap}{\alpha^{\prime}}
\newcommand{\be}{\begin{equation}}
\newcommand{\ee}{\end{equation}}
\newcommand{\ba}{\begin{eqnarray}}
\newcommand{\ea}{\end{eqnarray}}

\begin{document}
\hbox{\hskip 12cm ROM2F-97-6  \hfil}
\begin{center}  

{\large  \bf  TYPE I \ SUPERSTRINGS \ WITHOUT \ D-BRANES}
\vspace{0.5cm}

{\large  Massimo BIANCHI }
\vspace{0.3cm}

{\sl Dipartimento di Fisica, \ \ Universit{\`a} di Roma \ ``Tor Vergata''
\\ I.N.F.N.\ - \ Sezione di Roma \ ``Tor Vergata'', \ \ Via della Ricerca 
Scientifica, 1
\\ 00133 \ Roma, \ \ ITALY}
\vspace{0.5cm}
\end{center}
\begin{abstract} 
{\small Notwithstanding the central role of D-branes in many
recently proposed  string dualities, several interesting type I vacua have  been
found without  resorting directly to D-brane technology. In this talk, we
analyze a three-generation $SO(8)\times U(12)$ chiral type I model with $N=1$
supersymmetry in $D=4$. It descends from  the type IIB compactification on the 
Z orbifold and requires only Neumann boundary conditions, {\it i.e.} only the
ubiquitous D9-branes (pan-branes).  We also discuss a large class of 6D type I
vacua that display rich patterns of Chan-Paton symmetry breaking/enhancement and
various numbers of tensor multiplets.  Finally, we briefly address issues raised
by the conjectured  heterotic - type I duality and by the relation between type
I vacua and  compactificationsof the putative F-theory.} \end{abstract}

\vspace{0.5cm}

\section {Superstring Dualities: a Motivation}

Thanks to their soft ``Regge behaviour" at high energies, strings have been
regarded for some time as the only viable fundamental constituents of matter
\cite{gsw}. More recently, extended objects (p-branes) that had made their
appearance as string solitons
\cite{dkl} have suggested a more democratic attitude. The most fashionable
possibility is a theory of membranes (2-branes) and penta-branes (5-branes),
known as M-theory, that almost by definition has 11D supergravity as its
low-energy limit \cite{schw}.  This theory has no analogue of the dilaton, whose
(undetermined) vacuum expectation value plays the role of the string coupling
constant, and does not allow a perturbative expansion prior to compactification.
Upon dimensional reduction to 
$D=10$ M-theory gives the type IIA superstring \cite{witdyn}. Upon
compactification on a segment it is conjectured to give the $E(8)\times E(8)$
heterotic string
\cite{hw}.  A host of conjectured string dualities find a natural explaination
in terms of M-theory and of the putative  12D F-theory \cite{vafaf}. The latter
is related to the type IIB superstring after compactification on a two-torus. 
In the intricate web of conjectured dualities a central role is played by the 
type II solitons carrying Ramond-Ramond (R-R) charges. These have been formerly
identified with charged black p-brane solutions of the string equations of
motion to lowest order in  the inverse tension, $\ap$.  A microscopic description
\cite{pol} in terms of open strings with Dirichlet boundary conditions (=
D-branes) has opened the way to remarkable  progress not only in the string
duality realm but also in the context of black-hole thermodynamics \cite{sv}. 
Despite the enormous success of D-brane technology, this talk will mainly
present recent results obtained in the theory formerly known as type I
superstring. In order to keep it self-contained, we will start by recalling some
well-known facts about strings, p-branes and dualities. 

  At the perturbative level, there are two rather distinct classes of 
superstring theories: those with only closed oriented strings (type II A and B,
$E(8)\times E(8)$ and $SO(32)$ heterotic) and those with open and closed
unoriented strings (type I).  In the past, open-string models have been studied
to a lesser extent than models of oriented closed strings. Though the initial
proposal \cite{aug} of identifying open-string theories as {\it world-sheet
orbifolds} of left-right symmetric theories of oriented closed strings had
already been brought to a fully consistent systematization \cite{bs},
phenomenological considerations have favoured 4D perturbative vacua of the
$E(8)\times E(8)$ heterotic string preserving $N=1$  supersymmetry. After the
work of Seiberg and Witten on $N=2$ supersymmetric Yang-Mills theories \cite{sw},
these motivations seem to become much less compelling and a rather appealing
scenario is taking shape according to which different string theories should be
regarded as dual manifestations of a more fundamental (MF) theory \cite{sen}.
 
The best established of all string dualities is T-duality, a perturbative
symmetry between large and small volumes \cite{gpr}.  After compactification on
a circle of radius $R$, the presence of winding states in the closed-string
spectrum gives rise to a striking symmetry under 
$R\rightarrow \ap /R$. At the fixed point of the transformation new massless
states appear that enlarge the Kaluza-Klein $U(1)$ symmetry to $SU(2)$ through
the  Halpern - Frenkel - Kac (HFK) mechanism. In more complicated situations
T-duality becomes an infinite dimensional discrete symmetry  which acts on the
internal components of the metric $G_{ij}$ and antisymmetric tensor
$B_{ij}$. From the low-energy point of view these are scalar fields with only
derivative interactions that parametrize the space of classical vacua, known as
the  moduli space of the compactification.  From the world-sheet point of view
T-duality is a L-R asymmetric parity operation  on the string coordinates,
$\partial X \rightarrow
\partial X$ {\it but} $\bar\partial X
\rightarrow -\bar\partial X$, as well as on their superpartners. T-duality
interchanges type IIA and type IIB, the two heterotic theories (after the
introduction of Wilson lines), as well as Neumann with Dirichlet boundary
conditions for open strings. Thus, at the perturbative level T-duality  is not a
symmetry for type I superstrings \cite{bps}. Moreover, the asymmetric origin of
the scalar fields ($G_{ij}$ from the closed string NS-NS  sector, 
$B_{ij}$ from the  closed string R-R sector and the internal components of the
gauge fields  $A^a_i$ from the open string sector) prevents the existence of any
natural perturbative symmetry among them.  

In toroidal compactifications of the heterotic string to $D=4$, T-duality is
conjectured  to be accompanied by a non-perturbative symmetry called S-duality
\cite{ss}. In the $N=4$ low-energy effective lagrangian, the dilaton and the
axion parametrize the coset $SL(2,R)/U(1)$. The
$SL(2,R)$ symmetry of the lowest-order equations of motion is believed to be
broken by non-perturbative (space-time instanton) effects to $SL(2,Z)$. The
surviving symmetry, termed S-duality, includes transformations between strong and
weak coupling and generalizes the electric-magnetic duality of Maxwell theory. 
In type II theories, transformations between NS-NS and R-R states enlarge 
T-duality  to a symmetry known as U-duality \cite{ht}. In $D=4$ the discrete
subgroup 
$E(7,7;Z)$ of symmetries of the $N=8$ low-energy supergravity includes S-duality 
transformations and is conjectured to survive as a non-perturbative U-duality
symmetry of the full quantum theory. Turning off the gravitational interactions,
non-perturbative dualities in superstring  theory  may help explaining the
origin of some electric-magnetic dualities found in supersymmetric Yang-Mills
theories \cite{sw}.

\section{p-brane Democracy and Open-String Aristocracy}
 
Strong-weak coupling duality between charges and monopoles in $D=4$ admits a
natural  generalization to extended objects (p-branes) \cite{dkl}.  In $D$
dimensions the dual of the  $(p+2)$-form field strength of a
$(p+1)$-form potential is a $(D-p-2)$-form. This is the field strength of a
$(D-p-3)$-form potential naturally coupled to a $(D-p-4)$-brane.
Electric-magnetic duality in $D$ dimensions thus exchanges $p$-branes and
$(D-p-4)$-branes \cite{dkl}. As a corollary strings are believed to be dual to
penta-branes in $D=10$ and to strings in $D=6$.  In the latter case, a candidate
dual pair is provided by the heterotic string compactified on a four-torus  and
the  type IIA superstring compactified on a K3 surface \cite{ht,witdyn}.
 At points in the moduli space $SO(20,4)/SO(4)\times SO(20)$ where the abelian
gauge symmetry $U(1)^{24}$ is enhanced through the  HFK mechanism in the
heterotic string, one expects the appearence of new massless states in the type
IIA string. These states should be charged with respect to R-R  vector fields
and cannot belong to the perturbative spectrum.

Non-renormalization theorems in supersymmetric  theories, that guarantee the
stability of states saturating the Bogomolny - Prasad - Sommerfield (BPS) bound
between mass and central charges, suggest an interpretation in terms of BPS
solitons.  In $D=4$, these states are to be identified with charged extremal
black-holes 
\cite{ht}. From the higher (ten/eleven/twelve) dimensional point of view these
states may be pictured as p-branes wrapped around internal dimensions. When a 
homology cycle of the internal manifold shrinks to zero size, a conifold
transition may take place and the solitonic state whose mass is proportional to
the size of the cycle becomes massless \cite{strom}.  

Recently these ideas have attracted much attention in connection with the
observation that non-perturbative corrections in the string coupling $g$  may be
as large as $exp-{1\over g}$, rather than the expected $exp-{1\over g^2}$
\cite{pol}.  Since a boundary contributes one half a handle to the Euler
characteristic of the world-sheet, the open-string coupling constant is the
square-root of the closed string one. Thus, non-perturbative effects of the 
above kind are naturally generated by the  introduction of boundaries,
{\it{i.e.}} by coupling open strings to the closed string  spectrum.  Type II
theories admit solitonic p-brane solutions   which couple to R-R fields (with
odd p in the type IIB case and even p in the type IIA case) and whose masses
scale as
$M_{RR}\approx {1\over g}$.  A microscopic description is possible in terms of
open strings with some of the string  coordinates satisfying Dirichlet rather
than  Neumann boundary conditions, whence the name ``D-branes" \cite{pol}
\footnote{A R-R vertex operator with minimal, non-derivative, coupling to
boundary and crosscap states was found in \cite{bps}. But only  after the
seminal work of Polchinski \cite{pol}, the importance of the above coupling has
been fully appreciated.}. Under T-duality a  Dp-brane is exchanged with a
D(p$\pm$1)-brane, consistently with the interchange of the two type II theories.
Simple supersymmetric  configurations of D-branes correspond to
parallel/intesecting  Dp-branes  and D(p+4)-branes. The multiplicity of parallel
D-branes allows for a geometric interpretation of Chan-Paton (CP) factors. 

Open strings and D-branes play a crucial role in many recently conjectured
string  dualities \cite{sen}.  The $SO(32)$ type I superstring may be considered
as describing the excitations of a BPS configuration of 32 type IIB D9-branes.
Similarly, the excitations of the type I D-string (D1-brane) exactly coincide
with  the (light-cone) degrees of freedom of the
$SO(32)$ heterotic string \cite{pw}. Combined with the possibility of a direct
map between  the 10D low-energy effective actions, this observation has led to
conjecture a strong-weak coupling duality between the type I and heterotic 
$SO(32)$ theories. This conjecture has passed several tests \cite{tse} and is
expected to persist, in some non-naive form,  after compactification \cite{sen}.
In this respect, it is crucial to observe that the  relation between type I and
heterotic dilatons depends on the space-time dimension $D$ according to
\cite{chiral} 
\be
\phi_{H} = {6-D \over 4} \ \phi_{I} - {D-2 \over 16} \ \log \det G_I \quad ,
\ee where $G_I$ is the internal metric in the type I string-frame. For
instance,  in
$D=6$ \cite{blpssw}, the heterotic dilaton, that belongs to the universal tensor
multiplet, is related to the internal volume mode of the type I string, that 
belongs to a hypermultiplet \cite{bs, gepner}. 
 
In the unifying picture emerging from string dualities,  all p-branes should be
considered on an equal footing, whence the proposed {\it p-brane democracy} 
\cite{ht}.  Open strings, the best known  of all D-brane excitations, play
however a priviledged role - whence the proposed {\it open-string aristocracy} -
in that many  features that are non-perturbative in other descriptions are
reachable at a perturbative level in type I vacua
\cite{bs, gepner}.

\section{Type I Systematics}

The general construction of perturbative open-superstring vacuum configurations
consists in a non-standard $Z_2$-orbifold procedure \cite{aug, ps}
\footnote{Many authors refer to these {\it parameter space orbifolds} as {\it
orientifolds} though only a restricted class of the resulting models requires
orientifold planes.}. First of all, the conventional Polyakov perturbative series
must be supplemented with the inclusion  of world-sheets  with boundaries and/or
crosscaps. (Super)conformal field theories on  surfaces of this kind are
equivalent to (super)conformal field theories on  double-covering surfaces
endowed with a $Z_2$-orbifold projection of the spectrum under the exchange of
left-movers with right-movers. Roughly speaking, this procedure halves the
world-sheet symmetries  as well as their target space byproducts as expected for
BPS solitons. The truncation of the closed-string spectrum encoded in the torus
partition function ${\cal T}$ is implemented by the Klein bottle projection
${\cal K}$.  These two  contributions make up the ``untwisted sector" of the
parameter space orbifold
\be {\cal Z}_u = {1\over 2} ({\cal T} + {\cal K}) =  {1\over 2}
Tr_{closed}\left( (1+\Omega) q^H \bar{q}^{\bar{H}}\right)
\label{untwisted}
\ee  where $\Omega$ denotes orientation reversal. The role of the ``twisted
sector" is played by the open string spectrum  encoded in the annulus partition
function ${\cal A}$ and its projection, the M\"obius  strip~${\cal M}$ 
\be {\cal Z}_t = {1\over 2} ({\cal A}  + {\cal M}) =  {1\over 2} Tr_{open}\left(
(1+\Omega) q^H\right)
\label{twisted}
\ee In standard orbifolds, the twisted sectors have multiplicities associated  to
the fixed points.  Similarly, in parameter-space orbifolds, the open-string
states may acquire multiplicities associated to their ends through the
introduction of Chan-Paton (CP) factors. Consistency requirements  may be
deduced transforming the above amplitudes to the transverse channel, where Klein
bottle $\widetilde{\cal K}$, annulus 
$\widetilde{\cal A}$ and M\"obius strip 
$\widetilde{\cal M}$ are to be identified with closed-string amplitudes between
boundary  and/or  crosscap states. Since ``half" of the closed-string states
have been projected  out of the spectrum, it would be inconsistent if the latter
were to couple  to the vacuum. The cancellation between boundary and  crosscap
contributions to these tadpoles constrains the  CP factors and the signs in the
projections $\widetilde{\cal K}$ and $\widetilde{\cal M}$ \cite{pol,bs}. 

In order to explicitly solve these constraints, it proves very useful to exploit
Cardy's proposal and associate a boundary state $|B_i>$ as well as a CP factor
$n^i$ to each sector of the spectrum. This amounts to expressing the annulus
partition function as \cite{bs}
\be {\cal A}= \sum_{k} n^i n^j N_{ij}^k \chi_k 
\ee where $N_{ij}^k$ are the fusion rule coefficients  and the sum runs over the
sectors of the spectrum encoded in the characters $\chi_k$. In the transverse
channel,
$\widetilde{\cal A}$ becomes a linear combination of characters with
coefficients that are perfect squares thanks to the Verlinde's formula. Then
$\widetilde{\cal M}$ is fixed   for consistency as an appropriate ``square root"
of the product of
$\widetilde{\cal K}$ bottle and $\widetilde{\cal A}$. Following this procedure
for rational models a web of perturbatively consistent (non)supersymmetric type
I vacua may be constructed in any  dimension \cite{bs, mbas}. 

However, this is not the whole story.  Indeed the above construction assumes a
unique ${\cal K}$. Sewing constraints for conformal field theories on closed
oriented  Riemann surfaces can be extended to surfaces with boundaries and/or
crosscaps and a crosscap constraint can be deduced \cite{prad}.  In many
interesting cases  there are several solutions to this constraint which allow to
deduce several different projections of the closed string spectrum. The
procedure is then reversed, the number of allowed boundary states is  reduced
and may be inferred from $\widetilde{\cal K}$. Many new type I descendants of
type II B models  can be constructed systematically. In general, ${\cal K}$
allows for the introduction of signs in the projection of the closed-string
spectrum. For instance, in toroidal compactifications  it is possible to choose 
\be {\cal K} =
 \sum_{m\ {\rm even}} q^{ ({m\over R})^2} + \epsilon \sum_{m\ {\rm odd}} 
q^{({m\over R})^2}\quad , 
\label{exotic}
\ee where the conventional choice is $\epsilon = 1$. When $\epsilon=-1$,  there
are no massless tadpoles, and thus one cannot introduce  boundary states and
open strings 
\cite{gepner}.   This procedure may be generalized to any (rational) model
admitting $Z_2$ automorphisms, and the introduction of the open-string sector is
simply forbidden if $\widetilde{\cal K}$  does not involve massless characters.
More general toroidal compactifications were  discussed in \cite{bps}. There it
was shown how  a quantized background for the NS-NS antisymmetric tensor reduces
the size of the CP group. Deformations of the metric as well as of the gauge
fields are allowed and CP symmetry breaking may proceed via Wilson lines.

\section{D=6: Tensors and CP Symmetry Enhancement}

The first consistent 6D
$N=1$ open-string models \cite{bs, mbas} differ markedly from perturbative
heterotic K3 compactifications. The open-string spectra often  require
symplectic CP groups and the closed-string spectra contain different numbers of
(anti)self-dual tensors that take part in a generalized Green-Schwarz (GS)
mechanism
\cite{gss}.  Recently, additional instances of 6D $N=1$ type I models  have been
constructed as toroidal orbifolds \cite{gpdp}, along the lines of \cite{ps}. A
nice geometrical setting for all these 6D models is provided by the F-theory
proposal  of
\cite{vafaf}, where non-trivial scalar backgrounds allow for an effective 12D
dynamics. The variety of 6D models with different numbers of tensor multiplets
may  then be related to a corresponding variety of compactifications on 
elliptically-fibered CY threefolds
\cite{vafaf}. Many of these vacua are related to non-perturbative heterotic 
vacua with 5-branes in which the gauge symmetry is enhanced at the core of a 
Yang-Mills instanton \cite{blpssw}.

Exactly solvable K3 compactifications also arise from Gepner models and from
fermionic constructions. The starting point for a type I model is a ``parent''
type IIB theory, whose chiral spectrum is uniquely fixed by target-space
$N=(2,0)$ supersymmetry.  In the open descendants the surviving supersymmetry is
$N=(1,0)$, and the unoriented closed spectrum consists of the  supergravity
multiplet coupled to
$n^{c}_{T}$ tensor multiplets  and 
$n^{c}_{H}$ hypermultiplets. It should be appreciated that $n_{T}^{c} +
n^{c}_{H} = 21$ since all matter fermions arise from the 21 type IIB tensorini.
The open unoriented sector contains 
$n^{o}_{V}$ vector multiplets and $n^{o}_{H}$ charged  hypermultiplets. The
tensor fields that flow in the transverse channel take part in a generalized GS
mechanism
\cite{gss}.  The $U(1)$ anomalies are not cancelled by this GSS mechanism, but
involve the four-form dual of RR scalars, in a way reminiscent of the  Dine -
Seiberg - Witten  (DSW) mechanism in $D=4$ \cite{dsw}. 

Barring U(1) anomalies, the largest allowed CP group, $U(16)\times U(16)$,
requires the introduction of discrete Wilson lines \cite{mbas}, related to 
M\"obius projections that do not respect the accidental extended symmetry of the
rational model. After Higgsing, this model, which has been termed the GP
model\footnote{I suppose that  GP stand for the initials of Gianfranco Pradisi,
but I have not been able to trace the origin of this terminology in the
literature.}, 
 seems to be connected to perturbative vacua of the heterotic string in $D=6$
\cite{blpssw}. A large class of perturbative type I vacua with different  numbers
of tensor multiplets, including zero, and a rich pattern of symmetry
breaking/enhancement is available \cite{gepner} that extends the already rich
list of \cite{bs, mbas}. Not all of these type I vacua admit a clear D-brane
interpretation. It is tempting to conjecture that some of them should correspond
to configurations of D7-branes partially wrapped around the homology 2-cycles of
a K3 surface.

\section{A Chiral Type I Vacuum Configuration in $D=4$}

The first chiral $N=1$ supersymmetric type I model \cite{chiral} includes an
open-string sector with only Neumann boundary conditions, {\it i.e.} open
strings ending on the ubiquitous pan-brane. It has been found as the open
descendant of the type IIB  superstring compactified on the
$Z$ orbifold, the $Z_3$ orbifold of a six-torus.   The original massless
closed-string spectrum, containing the $N=2$ supergravity multiplet,
$10$ untwisted and 27  twisted  hypermultiplets, is truncated to $N=1$
supergravity coupled to $36$ chiral multiplets. Tadpole cancellations require
the introduction of open-string states with a CP gauge group 
$SO(8) \times SU(12) \times U(1)$. The resulting massless spectrum  includes
three generations  of chiral multiplets in the $(8,12^*)$ and $(1,66)$
representations \cite{chiral}. The
$U(1)$ factor is anomalous, and the DSW mechanism involving R-R axions gives a
mass of the order of the string scale to the corresponding gauge boson
\cite{dsw}. It is worth noticing that a candidate dual heterotic model is  a $Z$
orbifold with non standard embedding. In the twisted heterotic spectrum massless
states  in the spinor of $SO(8)$ appear which have no type I counteparts. In
order to put heterotic - type I $N=1$ duality on a firmer basis it is rather
compelling to understand the origin of these  states and to extend the list of
$N=1$ dual pairs. 

Returning to the untwisted sector of the $Z$ orbifold, the ``parent'' type IIB
model includes 20 NS-NS fields 
$(\phi, b_{\mu \nu}, b_{i \bar j}, g_{i \bar j})$, and 20 R-R fields   
$(\phi^{\prime}, b^{\prime}_{\mu \nu}, b^{\prime}_{i \bar j},  A_{\mu \nu i \bar
j})$. These parametrize the quaternionic manifold
$E_{6(+2)} /(SU(2) \times SU(6))$, obtained by c-map 
\cite{cfg} from the special K\"ahler manifold $SU(3,3)/(SU(3) \times SU(3)
\times U(1))$  of the  heterotic string.   In the NS-NS sector, the type I
projection retains the dilaton and a 9-dimensional real slice of the complex
K\"ahler cone corresponding to $Im(J_{i \bar j}) \ = \ Re(g_{i \bar j}) +
Im(b_{i \bar j})$. In the R-R sector, one is left with a mixture of
$\phi^{\prime}$ and 
$b^{\prime}_{\mu \nu}$, as well as with mixtures of the other R-R fields.  
Though somewhat surprising, this result is clearly encoded in the partition
function  and is also supported by the explicit study of tree-level amplitudes
\cite{twelve}. The 20 scalar fields parametrize a space  ${\cal S}_I$  that, on
general supersymmetry grounds, is a K\"ahler manifold  embedded in
$E_{6(+2)} /(SU(2) \times SU(6))$. Group theory considerations together with
some rather compelling duality arguments uniquely select 
${\cal S}_I = Sp(8,R)/(SU(4) \times U(1))$ \cite{twelve}, a generalized Siegel
upper half plane. It should be appreciated that  the type I truncation is vastly
different from the heterotic one. Observe that ${\cal S}_I$ is the space of
symmetric $4\times 4$ matrices $\Omega$ with positive $Im\Omega$ and has a
natural K\"ahler metric with K\"ahler potential
$K = - \log \det Im \Omega $. The elements of $Re\Omega$ are R-R scalars, while 
the elements of
$Im\Omega$ parametrize a real slice of the complex K\"ahler cone of the CY
threefold. The symplectic group $Sp(8,R)$ acts naturally on the ``Teichm\"uller"
space ${\cal S}$ through projective transformations of
$\Omega$ and  clearly includes $S$ and $T$ duality transformations. In
particular, the  continuous Peccei-Quinn (PQ) symmetry of the R-R fields
corresponds to the subgroup of $Sp(8,R)$ triangular matrices
\be  S_{PQ} = \pmatrix{1 & B \cr 0 & 1 \cr} \quad ,
\ee  with $B=B^T$. It is natural to expect that (world-sheet and space-time)
instanton effects induce non-trivial monodromies with respect to 
$Sp(8,Z)$, so that the moduli space is actually ${\cal S}_I/Sp(8,Z)$.  A 
non-perturbatively generated superpotential would then be a modular form of
$Sp(8,Z)$.  

The above results suggest that open descendants of type IIB compactifications on
generic  CY threefolds involve a new complexification of the classical moduli
space. The new map for open descendants, termed ``o-map"
\cite{twelve}, associates a type II quaternionic  manifold ${\cal Q}_{n+1}$ to a
new K\"ahler manifold $K_I$ for $(n+1)$ NS-NS and
$(n+1)$ R-R real scalars. The complexification of the K\"ahler class of
classical CY threefolds is at the heart of mirror symmetry, that should also
have a new realization in the type I setting. In this case the $N=1$ classical
K\"ahler manifold should  receive quantum corrections from D-brane world-volume
instantons \cite{bbs}.

\section{Glimpses of the F-Theory}

Non-trivial backgrounds for the scalar fields of type IIB supergravity have
provided a geometrical setting \cite{vafaf} for some peculiar  6D string vacua
previously derived as open descendants of type IIB K3 compactifications
\cite{bs, mbas}.  The resulting models differ markedly  from conventional K3
reductions, since their massless spectra contain different numbers of
(anti)self-dual tensors that take part in the GSS mechanism
\cite{gss}. These peculiar features find an elegant rationale in the
compactification of a putative 12D F-theory \cite{vafaf} on elliptically fibered
Calabi-Yau (CY) threefolds, a construction that generalizes previous work on 
supergravity vacua with scalar backgrounds by taking into account subtle global
issues. 

The simplicity of the type I model of
\cite{chiral} reflects itself in the linearly realized symmetry,
$H = SU(4) \times U(1)$. The $SU(4)$ factor strongly suggests a 12D
interpretation \cite{twelve} as the holonomy of a CY four-fold. The possible
relation of the type I superstring on the 
$Z$ orbifold  to  the F-theory reduction on a CY fourfold \cite{twelve} can only
give a hint to the  richness of open-string constructions.  The idea that a
fundamental non-perturbative description could be democratic with respect to the
various p-branes is rather appealing. Still perturbative computability tends to
favour one-branes. The recent results seem to indicate that open one-branes,
with all possible boundary conditions, are more equal than the others.

\section{Acknowledgements}

It is a pleasure to thank the organizers of the XII SIGRAV Meeting for their kind
inviation. I also wish to thank  C. Angelantonj, S. Ferrara, G. Pradisi, A.
Sagnotti, and Y. Stanev for a pleasant and fruitful collaboration on most of the
topics  presented in this talk.

\end{document}